\documentclass{article}
\usepackage{amsmath}
\usepackage{amssymb}
\usepackage{graphicx}
\begin{document}
\begin{center}
\Large
\textbf{Continuous symmetry entails the Jordan algebra structure of finite-dimensional quantum theory}
\vspace{0,3 cm}
\normalsize

Gerd Niestegge 
\footnotesize
\vspace{0,2 cm}

Ahaus, Germany

gerd.niestegge@web.de, https://orcid.org/0000-0002-3405-9356
\end{center}
\normalsize
\begin{abstract}
Symmetry postulates play a crucial role 
in various approaches to reconstruct quantum theory 
from a few basic principles. Discrete and 
continuous symmetries are under consideration. 
The continuous case better matches the physical needs for 
mathematical models of dynamical processes and is studied here.
Applying the representation theory of the orthomodular lattices
and a generalized version of Gleason's theorem for 
Jordan matrix algebras, we show that 
the continuous symmetry, together with three further requirements, 
entails that the underlying mathematical structure 
of a finite-dimensional generalized probabilistic theory becomes 
a simple Euclidean Jordan algebra. The further requirements are: 
spectrality, a strong state space and a condition called gbit property.
\vspace{0,3 cm}

\end{abstract}

\section{Introduction}

The \emph{generalized probabilistic theories} (GPTs \cite{barnum2014higher, barrett2007information, 
Chiribella-PRA2011, muller2021probabilistic}) are commonly used as generic models 
to gain a better understanding of the probabilistic and information theoretic 
foundations of quantum physics and quantum computing or
to reconstruct quantum theory from a few basic principles.
In the various reconstructions \cite{muller2021probabilistic},
symmetry postulates play a crucial role. 
In many cases \cite{BarnumHilgert2020, barnum2014higher, masanes2011derivation, muller2012ududec}, 
the symmetries under consideration are discrete. Continuous symmetries would
better match the physical needs for mathematical models of dynamical
processes and are considered in other approaches \cite{hardy2001from5axioms}.

So do we here and study the impact of the continuous symmetry
on the underlying mathematical structure of the finite-dimensional GPTs. 
We show that this becomes a simple (irreducible) 
\emph{Euclidean Jordan algebra} \cite{AS02, faraut1994analysis, hanche1984jordan},
when the continuous symmetry, spectrality, a strong state space and one further 
condition are given. Our proof is based on the representation 
theory of the orthomodular lattices \cite{maclaren1965notes, piron1964axiomatique}. This requires the
\emph{covering property}, which we shall derive from the more information theoretical 
\emph{gbit-property} - the further condition. Furthermore, we need a generalized version  
of Gleason's theorem for Jordan matrix algebras \cite{bunce1985Jordan}.
\newpage

The GPTs usually result in order unit spaces,
which become out basic structure. In section~2 we 
provide a motivation for it, which is rather independent of the GPTs,
and introduce the concept of spectrality.
Strongly order determining state spaces and the consequences
are studied in section~3.
Irreducibility is considered in section~4, the gbit-property in section~5.
The symmetry postulates are presented in section~6,
before the main results are elaborated in section~7.

\section{Spectral order unit spaces}

In quantum theory, self-adjoint linear operators represent the observables.
The usual operator product does not remain self-adjoint, only the Jordan product does,
but the physical meaning of the Jordan product of non-commuting observables is not clear.
Therefore the question arises what would become an appropriate mathematical structure 
for the system of observables, when we do not insist on the existence of a product
from the beginning.

Any linear space of self-adjoint linear operators that contains the identity operator $\mathbb{I}$
and any linear space of real functions that contains the constant function $\mathbb{I} \equiv 1$
forms an \emph{order unit space} \cite{AS02} with the usual order relation, the operator norm in the first case
and the sup norm in the second case. Vice versa, any order unit space can be represented as a
linear space of real functions (on a certain set) with the sup norm.
An excellent candidate for the desired structure for the observables 
thus becomes an order unit space $A$; this means there are an order unit $\mathbb{I} \in A$ 
and a norm $\left\|\ \right\|$ such that 
$ \left\|a \right\| = inf \left\{s \in \mathbb{R} | -s \mathbb{I} \leq a \leq s \mathbb{I} \right\}$
for $a \in A$. 

In quantum theory, the projection operators $p = p^{2}$ form 
the so-called \emph{quantum logic} \cite{maclaren1964atomic, maclaren1965notes, 
piron1964axiomatique, zierler1961axioms, zierler1966lattice}
and play a special role in the spectral resolutions of the observables.
The spectral resolution is an important feature of the 
mathematical formalism of quantum theory and is needed to define the potential numerical measurement outcomes 
of an observable (its spectrum) and the probability distributions over them.
Without the product we cannot define the projections via $p^{2} = p$. However, there is 
another characterization of the projections in the self-adjoint part of a von Neumann algebra $M$; the projections
are the extreme points of the convex set $\left\{a \in M| 0 \leq a \leq \mathbb{I} \right\}$ \cite{AS02}.
This characterization remains available in any order unit space $A$ and we consider 
$$L_A := ext \left[0,\mathbb{I}\right]$$
with $ \left[0,\mathbb{I}\right] := \left\{a \in A | 0 \leq a \leq \mathbb{I}\right\}$
as candidate for the quantum logic. Note that $0, \mathbb{I} \in L_A$ and 
$p \in L_A \Leftrightarrow \mathbb{I} - p \in L_A$. Thus we can define the following
\emph{orthocomplementation} $'$ on $L_A$: $p' := \mathbb{I} - p$. Two elements $p$ and $q$ in $L_A$
with $q \leq p'$ or equivalently $p + q \leq \mathbb{I}$ are called \emph{orthogonal}.
Note that all non-zero elements in $L_A$ carry the norm $1$.
\newpage

Generally, it is not clear
that $L_A$ contains more elements than only $0$ and $\mathbb{I}$. When
$A$ has a finite dimension, $\left[0,\mathbb{I}\right]$ is the 
convex hull of its extreme points by Minkowski's theorem and $L_A$
become sufficiently large. Moreover, with the affine transformation 
$T(a) = 2a - \mathbb{I}$, $a \in A$, we get 
\begin{align*}
\left\{a \in A \;|\; \left\| a \right\| \leq 1\right\} 
&= \left\{a \in A \;|\; - \mathbb{I} \leq a \leq \mathbb{I}\right\}\\
&= T(\left[0,\mathbb{I}\right]) 
 = conv(\left\{2p -\mathbb{I} \;|\; p \in L_A \right\})\\
&= conv(\left\{p -p' \;|\; p \in L_A \right\}).
\end{align*}
For any 
linear functional $\rho : A \rightarrow \mathbb{R}$ we then have
\begin{align*}
\left\| \rho \right\| 
&= sup \left\{\left|\rho(a)\right| : \left\|a\right\| \leq 1\right\}
 = sup \left\{\left|\rho( p ) - \rho (p') \right| : p \in L_A\right\}\\
&\leq 2 sup \left\{\left|\rho(q)\right| : q \in L_A\right\}.
\end{align*}

An \emph{atom} $e$ in this quantum logic is a minimal non-zero element in $L_A$; this means:
$ q \in L_A$, $0 \neq q \leq e \Rightarrow q = e$. The \emph{information capacity} $m$ of $A$ or $L_A$ 
is the maximum number of pairwise orthogonal non-zero elements in $L_A$.
Since such elements are linearly independet, $m$ is finite if the dimension of $A$ is finite,
but the dimension of $A$ need not be finite if $m$ is finite.

An order unit space $A$ with finite information capacity $m < \infty$ 
is here called \emph{spectral}, if each element $a \in A$ can be written as
$a = \sum s_k e_k$ with atoms $e_1, ..., e_n \in L_A$,  $s_1, ..., s_n \in \mathbb{R}$, $n \leq m$, 
and $ \sum e_k = \mathbb{I}$. In this form, the spectral representation is not unique
and therefore not sufficient to define the probability distribution of the observable $a$;
the uniqueness requires an additional assumption, which we consider in the next section. 

Typical examples of spectral order unit spaces are the Euclidean Jordan algebras, 
which are the same as the formally real Jordan algebras and the finite-dimensional 
JB- or JBW-algebras \cite{AS02, faraut1994analysis, hanche1984jordan}. They include the 
self-adjoint Jordan matrix algebras over $\mathbb{R}$ and $\mathbb{C}$.

\section{Strongly order determining state spaces}

A \emph{state} is a non-negative linear functional $\mu$ on $A$ with $\mu(\mathbb{I})=1$.
The state space is denoted by $S_A$. It is called \emph{strongly order determining} if
$$\left\{\mu \in S_A | \mu(p) = 1 \right\} \subseteq \left\{\mu \in S_A | \mu(q) = 1 \right\} \Rightarrow p \leq q$$
holds for the $p,q \in L_A$.

In any order unit space there is a state $\mu$ with $\mu(a) = \left\|a\right\| $ for each $a \in A$. 
Particularly for each $e_j$ in the above spectral resolution of $a \in A$ 
there is a state $\mu_j$ with $\mu_j(e_j)=1$; then $\mu_j(e_k)=0$ for $k \neq j$ and $\mu_j(a)= s_j$. 
Therefore $s_k \geq 0$ for each $k$, if $a \geq 0 $, and $0 \leq s_k \leq 1$ for each $k$, 
if $0 \leq a \leq \mathbb{I} $.

We shall now see that $L_A = ext \left[0,\mathbb{I}\right]$ becomes a convenient quantum logic 
and identify a unique form of the the spectral representation,
when the order unit space $A$ is spectral and its state space is 
strongly order determining.\\

\textbf{Lemma 1:} \textit{Let $A$ be a spectral order unit space with finite information capacity~$m$ 
and strongly order determining state space $S_A$}.
\begin{itemize}
\item[(i)]
$L_A$ \textit{is an orthomodular lattice.}
\item[(ii)]
\textit{Each element $a\in A$ has a unique spectral representation $a = \sum^{n}_{k=1} s_k p_k$
with pairwise orthogonal elements $p_k \in L_A$ and $s_k \in \mathbb{R}$ 
with $s_i \neq s_j$ for $i \neq j$. Then $n \leq m$,
$ \left\|a\right\| = max \left\{\left|s_k\right| : k=1,...,n \right\}$ 
and $a \geq 0$ iff $s_k \geq 0$ for each $k$.}
\end{itemize}

Proof. (i) We first show that $p - q$ lies in $L_A$, when $q \leq p$ holds with any $p,q \in L_A$. 
We have to verify that $p - q$ is an extreme point of $\left[0,\mathbb{I}\right]$.
Thus suppose $p - q = s a + (1-s) b$ with $0 < s < 1$ and $a,b \in \left[0,\mathbb{I}\right]$.
For any state $\mu$ with $\mu(q) = 1$ we get $\mu(p) = 1$ and 
$ 0 = \mu(p - q) = s \mu(a) + (1-s) \mu(b)$, thus $0 = \mu(a) = \mu(b)$ 
and $1 = \mu(\mathbb{I} - a) = \mu(\mathbb{I} - b)$. Because $S_A$ is strongly order determining,
we have $q \leq \mathbb{I} - a$ and $q \leq \mathbb{I} - b$ or, equivalently, 
$ a + q \leq \mathbb{I}$ and $ b + q \leq \mathbb{I}$.
Since $p$ is an extreme point, we get from $p = s (a + q) + (1-s) (b + q) $
that $a + q = p = b + q$. Therefore we have $p - q = a = b$ and
we have verified that $p - q$ is an extreme point.

Now suppose $p$ and $q$ are orthogonal in $L_A$. Then $ q \leq \mathbb{I} - p$
and $\mathbb{I} - p - q \in L_A$. Hence $p + q = (\mathbb{I} - p - q)' \in L_A$.
Particularly any sum of orthogonal atoms $e_1 + ... + e_n$ becomes an element in $L_A$.

If $q \in L_A$ is not an atom, there is an atom $e_1$ with $e_1 \leq q$. 
If $q - e_1 \in L_A$ is not an atom, there is a further atom $e_2$ with $e_2 \leq q - e_1$.
Continuing this process, we finally get $q = e_1 + ... + e_n$ 
with certain atoms $e_1, ..., e_n$, since this process stops after at most 
$m$ steps, where $m$ is the information capacity. Any element in $L_A$ thus becomes
the sum of atoms. 

We now consider two elements $p,q \in L_A$ and a spectral representation of their sum 
$p + q = \sum s_k e_k$; $e_1, ..., e_n$ are atoms with $\sum e_k = \mathbb{I}$. 
Then $0 \leq s_k \leq 2$ for each~$k$.
We first assume $s_k < 2$ for all $k$. If $0 \neq y \in L_A$ with $y \leq p$ and $y \leq q$,
there is $\mu \in S_A$ with $\mu(y)=1$. We have $2y \leq p + q$. Thus we get the contradiction
$2 \leq \mu(p+q) = s_1 \mu(e_1) + ... + s_n \mu(e_n) \leq max\left\{s_1, ..., s_n\right\} < 2$.
Therefore we have $p \wedge q = 0$.
We now assume $s_k = 2$ for some $k$ and define $x := \Sigma_{k:s_k=2} \ e_k \in L_A$.
Suppose $\mu(x) = 1$ with $\mu \in S_A$. The spectral representation then 
gives $ \mu(p+q) = 2$. Since $\mu(p) \leq 1$ and  $\mu(q) \leq 1$,
we get $ \mu(p) = 1 = \mu(q)$. 
Therefore $x \leq p$ and $x \leq q$ because $S_A$ is strongly order determining.
Now suppose $y \in L_A$ with $y \leq p$ and $y \leq q$.
For any $\mu \in S_A$ with $\mu(y)=1$ we get $\mu(p) =1 = \mu(q)$ and $ \mu(p+q) = 2$.
The spectral representation gives $\mu(x)=1$. Therefore $y \leq x$
and we finally have $ x = p \wedge q $.
The supremum $p \vee q$ exists in $L_A$ since it is identical with $(p'\wedge q')'$
(see \cite{gudder1966uniqueness} for a very similar proof of the lattice properties).

Suppose again that $p$ and $q$ are orthogonal in $L_A$. 
Then $p = e_1 + ... + e_k$ and $q = e_{k+1} + ... + e_l$
with atoms $e_1, ..., e_l$. 
Moreover $\mathbb{I} - e_1 - ... - e_l = e_{l+1} + ... + e_n$ with $n \leq m$.
Then $p' + q' = e_1 + ... + e_l + 2 (e_{l+1} + ... + e_n)$ and
$p' \wedge q' = e_{l+1} + ... + e_n $ (see above). Thus 
$p \vee q = (p' \wedge q')' = e_1 + ... + e_l = p + q$.
\newpage

Now we consider $p,q \in L_A$ with $q \leq p$ and show $p - q = p \wedge q'$. 
We have $p - q = p \in L_A$.
Obviously $p - q \leq p$, $p - q \leq \mathbb{I} - q$ and therefore $ p \wedge q' \leq p - q$.
If $x \in L_A $ with $p \leq x$ and $q' \leq x$, then $p - q \leq p \leq x$
and $p - q \leq \mathbb{I} - q = q' \leq x$ and therefore $p - q = p \wedge q'$.

Finally $p = q + (p-q) = q + p \wedge q' = q \vee (p \wedge q')$ and we have verified the 
orthomodular law. 

(ii) The existence of such a representation immediately follows from the
spectrality by using the sums of those atoms $e_k$ for which the $s_k$ coincide.

Now suppose $a = \sum^{n}_{k=1} s_k p_k$ as stated above. After some renumbering, we
can assume that $s_1 > s_2 > ... > s_n$. Then 
$s_1 = max\left\{\mu(a) | \mu \in S_A\right\}$ and
$\mu(a) = s_1$ iff $\mu(p_1) = 1$.
With a second representation $a = \sum t_k q_k$
with pairwise different and ordered $t_k$, we get
$t_1 = max\left\{\mu(a) | \mu \in S_A\right\} = s_1$
and $\mu(q_1) = 1$ iff $\mu(a) = t_1 (= s_1)$ iff $\mu(p_1) = 1$.
Since the state space is strongly order determining, we have $p_1 = q_1$.

In the next step consider 
$a_1 = a - s_1 p_1 = a - t_1 q_1 = \sum_{k \geq 2} s_k p_k = \sum_{k \geq 2} t_k q_k$
and repeat the above procedure
to get $s_2 = t_2$ and $p_2 = q_2$. Continuing this process finally
shows the uniqueness of the above form of spectral representation after $m$ steps
at most. 

We have $a \geq 0$ iff $\mu(a) \geq 0$ for all $\mu \in S_A$
and this is equivalent to $s_k \geq 0$ for each $k$,
since there is a state $\mu$ with $\mu(p_k)=1$ each $k$
and then $\mu(p_j) = 0$ each $j \neq k $.
Moreover $ \left\| a \right\| 
= sup \left\{\left|\mu(a)\right| : \mu \in S_A \right\}
= max \left\{\left|s_k\right| : k=1,...,n \right\} $.
\hfill $\square$ \\

Note that, with the finite information capacity, 
the orthomodular lattice $L_A$ becomes also atomic and complete.

With the above spectral representation $a = \sum s_k p_k$, 
it suggests itself to define $a^{2} = \sum s_k^{2} p_k$, $n \in \mathbb{N}$.
If there were a product $\circ$ in $A$ with $a^{2} = a \circ a$,
it would immediately follow from a result by Iochum and Loupias \cite{IochumLoupias1985}
that $A$ becomes a Euclidean Jordan algebra.
A more elementary proof consists of the following three steps.
First note that $p \circ q = 0$ must hold with the above product 
for any orthogonal pair $p,q \in L_A$ . From \cite{Schafer1966} Lemma 5.2
we then get $p \circ (q \circ a) = q \circ (p \circ a) $
and the Jordan identity finally follows from the spectral representation.
However we want to avoid the assumption that the product exists 
and have to go another more laborious and longer way.

\section{Irreducibility}

With an order unit space $A$ and any $p \in L_A$,
let $A_p$ denote the set of those elements $a \in A$ for which there is 
a non-negative real number $s$ with $-s p \leq a \leq s p$; $A_p$ is
an order unit space with the order unit $p$.\\

\textbf{Lemma 2:} 
\itshape 
Let $A$ be a spectral order unit space with finite information capacity~$m$ and
a strongly order determining state space $S_A$. 

Each $a \in A_p$ with $p \in L_A$
has a spectral representation $a = \sum s_k e_k$ with $e_k \leq p$ 
for each $k$ with $s_k \neq 0$. 

Moreover, $L_{A_p} = ext(\left[0,p\right])$ coincides 
with $\left\{q \in L_A | q \leq p\right\}$ and $A_p$ is spectral.
\normalfont \\

Proof. Assume $0 \leq a \leq r p$ with some $p \in L_A$ and $0 < r \in \mathbb{R}$
and consider the spectral representation $a = \sum s_k e_k$, where $0 \leq s_k$ 
for each $k$ and the $e_k$ are atoms with $\sum e_k = \mathbb{I}$.
Let $\mu \in S_A$ be any state with $\mu(p')=1$.
Then $\mu(p)=0 $ and $0 = \mu(a) = \sum s_k \mu(e_k)$. Therefore
$\mu(e_k) = 0$ and $\mu(e_k') = 1$ for each $k$ with $s_k \neq 0$.
Since $S_A$ is strongly order determining, we thus have 
$p' \leq e_k' $ and $e_k \leq p$ for each $k$ with $s_k \neq 0$.

If $- r p \leq a \leq r p$, then $a + r p = \sum s_k e_k $ with $e_k \leq p$ for $s_k \neq 0$. 
With $p_o := \sum_{s_k \neq 0} e_k$ we have $p_o \leq p$ and 
$a = \sum s_k e_k - r p = \sum s_k e_k - r p_o  - r(p - p_o) 
= \sum_{s_k \neq 0} (s_k - r) e_k - r(p - p_o) $. Represent $p - p_o$ 
as a sum of atoms to complete the proof of the first part of the lemma.

Now suppose $x \in ext(\left[0,p\right])$. Then $0 \leq x \leq p$ and $x=\sum s_j e_j $ 
with $0 < s_j$ and $e_j \leq p$. If $s_{j_o} \neq 1$ for some $j_o$, we have
$x = s_{j_o} (e_{j_o} + \sum_{j \neq j_o} s_j e_j) + $ $ (1 - s_{j_o}) (\sum_{j \neq j_o} s_j e_j)$
and $x$ is not extremal. Thus we have $x = \sum e_j \in L_A$ 
Vice versa, if $x \in L_A$ with $x \leq p$, $x$ becomes extremal in $\left[0,p\right]$.
\hfill $\square$ \\

The state space
$S_{A_p}$ consists of the restrictions to $A_p$ of the states $\mu \in S_A$ 
with $\mu(p) = 1$ and is strongly order determining.

Two elements $p_1$ and $p_2$ in any orthomodular lattice $L$ are called \emph{compatible}, 
if $L$ contains three orthogonal elements
$q_1, q_2, q_3$ with $p_1 = q_1 + q_2$ and $p_2 = q_2 + q_3$; then $q_2 = p_1 \wedge p_2$. 
The \emph{center} is the set of all those elements
that are compatible with each element in $L$, and $L$ is called
\emph{irreducible} if the center is trivial, which means that the center is $\left\{0,\mathbb{I}\right\}$. 
Generally, the center is a Boolean sublattice, its minimal non-zero elements are pairwise orthogonal
and their sum amounts to $\mathbb{I}$.
Note that these elements are minimal in the center, but not in~$L$.
Each atom in $L$ lies under one of the minimal central non-zero elements.\\

\textbf{Lemma 3:} \textit{
Suppose $A$ is a spectral order unit space with finite information capacity $m$ and 
a strongly order determining state space $S_A$.
Let $c_1, ... c_n$ be the minimal non-zero elements in the center of $L_A$. Then $n \leq m$ and 
$A = A_{c_1} \oplus A_{c_2} \oplus ... \oplus A_{c_n}$}.\\

\textbf{Proof.} Each $a \in A$ has a spectral representation 
$a = \sum s_k e_k$ with atoms $e_1, ..., e_n \in L_A$, $s_1, ..., s_n \in \mathbb{R}$ 
and $ \sum e_k = \mathbb{I}$. For each $e_k$ there is a $c_j$ with $e_k \leq c_j$.
Therefore we have $a \in A_{c_1} \oplus A_{c_2} \oplus ... \oplus A_{c_n}$. 
\hfill $\square$ \\

\textbf{Lemma 4:} \itshape Suppose that $A$ is a spectral order unit space with
the finite information capacity $m$ and a strongly order determining state space.
If the atoms in $A$ form a connected set, then $L_A$ is irreducible.\\
\normalfont

\textbf{Proof.} Suppose that $L_A$ is not irreducible. Then there are at least two minimal non-zero elements $c_1$ and $c_2$ 
in the center and we select atoms $e_1$ and $e_2$ in $L_A$ such that $e_1 \leq c_1$ and $e_2 \leq c_2$.
Let $e_t, t \in \left[0,1\right]$, be a continuous path from $e_1$ to $e_2$
and define $s := inf \left\{ t: e_t \leq c_2 \right\} $.
The continuity then implies $e_s \leq c_2$ and thus $0 < s$. 
Note that the set $\left\{a \in A | 0 \leq a \leq p\right\}$ is closed for any $p \in L_A$.
For $t < s$ $e_t$ must lie under another minimal central non-zero element 
and we thus have $e_t \leq \mathbb{I} - c_2$.
The continuity implies $e_s \leq \mathbb{I} - c_2$. We then get 
the contradiction $0 \neq e_s \leq c_2 \wedge c^{'}_{2} = 0$.
\hfill $\square$

\section{The gbit property}

Spaces with the information capacity $m=2$ 
can be considered generalized models of a bit \cite{Nie2022genqubit}. 
We would expect that any pair of atoms generates such a model.
This motivates the following postulate for a spectral order unit space $A$
with finite information capacity and strongly order determiningg state space:\\

\textit{For any two atoms $e_1 \neq e_2$, $A_{e_1 \vee e_2}$ has the information capacity m = 2.}\\

This condition shall here be called \emph{gbit-property}. It can be stated in the following equivalent way:\\

\textit{If $e_1$ and $e_2$ are atoms and $ 0 \neq q \leq e_1 \vee e_2$ holds with $q \in L_A$,
then either $q = e_1 \vee e_2$ or $q$ is an atom}.\\

We shall now see that the gbit property is nothing else as the \emph{covering property} in disguise.
The covering property was studied a lot in lattice theory
\cite{AS01, AS02, piron1964axiomatique} and looks as follows:\\

\textit{If $e$ is an atom and $p \leq q \leq p \vee e $ for any $p,q$ in the lattice, then
$q = p$ or $q = p \vee e$.}\\

If the covering property holds in an orthomodular lattice, there is a dimension function $dim$
such that each element $p$ is the supremum
of $dim(p)$ orthogonal elements in $L$ \cite{AS01}.\\

\textbf{Lemma 5:} \textit{The following two conditions 
are equivalent for any orthomodular lattice $L$:}
\begin{itemize}
	\item [(i)] 
\textit{If $ 0 \neq q \leq e_1 \vee e_2$ with $q,e_1,e_2 \in L$,
where $e_1$ and $e_2$ are atoms, then either $q = e_1 \vee e_2$ or $q$ is an atom} (gbit-property).
	\item [(ii)] 
$L$ \textit{has the covering property}.
\end{itemize}
\textbf{Proof.}
(i) $\Rightarrow$ (ii): 
The proof becomes identical with the second part of the proof of \cite{AS02} Proposition 9.11.
The lattice of \emph{projective units} is considered there,
but the line of reasoning works for any orthomodular lattice.

(ii) $\Rightarrow$ (i): 
Suppose $L$ has the covering property and 
assume $ 0 \neq q \leq e_1 \vee e_2$ with $q,e_1,e_2 \in L$,
where $e_1$ and $e_2$ are atoms. 
Then $1 = dim(q)$ and $q$ is an atom,
or $2 \leq dim(q) \leq dim(e_1 \vee e_2) \leq dim(e_1) + dim(e_2) = 2$; 
in the second case we have $dim(q) = dim(e_1 \vee e_2)$ and then $q = e_1 \vee e_2$.
\hfill $\square$
\newpage

\section{Weak and continuous symmetry}

An automorphism of an order unit space $A$ is a an invertible positive transformation 
$T: A \rightarrow A$ with a positive inverse and $T(\mathbb{I}) = \mathbb{I}$. The 
automorphisms form a group denoted by $Aut(A)$. This becomes a compact group, 
if the dimension of $A$ is finite. A spectral order unit space $A$ 
shall here be called \emph{continuously symmetric}, if the following condition holds:\\

\textit{For any pair of atoms $e_1$ and $e_2$ there is a continuous map 
$t \rightarrow T_t$ from the real unit interval
$\left[0,1\right]$ to $Aut(A)$ such that $T_0$ is the identity in $Aut(A)$ and $T_1(e_1)=e_2$}.\\

This means that any atom $e_1$ can be continuously transformed to any other atom~$e_2$ 
in a reversible way and is one of Hardy's five axioms \cite{hardy2001from5axioms}, 
although he assumes this for the pure states instead of atoms.
The above condition can be reformulated as follows:\\

\textit{For any pair of atoms $e_1$ and $e_2$ there is a transformation $T$ 
in the connected component of $Aut(A)$ that contains the identity of this group
with $T(e_1)=e_2$}.\\

We shall also use the following weaker non-continuous form of symmetry; $A$ 
shall be called \emph{weakly symmetric}, if this holds:\\

\textit{For any pair of atoms $e_1$ and $e_2$ there is $T \in Aut(A)$ with $T(e_1)=e_2$}.\\

Weak symmetry means that the automorphism group $Aut(A)$ acts transitively on the atoms.\\

\textbf{Lemma 6:} L\textit{et $A$ be a finite-dimensional spectral order unit space 
with a strongly order determining state space. If $A$ is weakly symmetric, then
$A$ possesses an $Aut(A)$-invariant inner product $\left\langle \ |\  \right\rangle_o$ with 
$\left\langle e|e\right\rangle_o = 1$ for each atom~$e$.}\\

Proof. Let $\left\langle  \ | \ \right\rangle$ be any inner product on $A$. 
Using the normalized Haar measure on $Aut(A)$,
we construct the further $Aut(A)$-invariant inner product
\begin{center}
$ \left\langle a|b\right\rangle_o := \int_{U \in Aut(A)} \left\langle Ua|Ub\right\rangle dU $
\end{center}
for $a,b \in A$. Since $Aut(A)$ acts transitively on the atoms, we have 
$0 \neq \left\langle p|p \right\rangle_o = \left\langle q|q \right\rangle_o$ 
for any two atoms $p$ and $q$ and we can normalize 
$\left\langle \ | \  \right\rangle_o$ in such a way that 
$\left\langle e|e \right\rangle_o = 1$ for every atom $e$.
\hfill $\square$
\newpage

\section{Main results}

We first prove a theorem with most general assumptions, from which 
our main result will then follow as a corollary.\\

\textbf{Theorem 1:} \itshape Suppose that $A$ is a spectral order unit space with
finite dimension, information capacity $m$ and a strongly order determining state space.

If $m=2$ and $A$ is weakly symmetric, then $A$ is a spin factor (the spin factors are the 
irreducible Euclidean Jordan algebras with $m = 2$).

If $m \geq 4$, weak symmetry and the gbit-property are given
and the atoms form a connected set, then $A$ is a 
simple (irreducible) Euclidean Jordan algebra and can be represented 
as the Jordan matrix algebra of Hermitian $m \times m$-matrices over the real numbers, 
complex numbers or quaternions.\\
\normalfont

Proof. (i) Suppose $m=2$ and $A$ is weakly symmetric. If $e_1$ and $e_2$ are atoms,
there is $T \in Aut(A)$ with $T(e_1) = e_2$. Then $T(e_1') = e_2'$ and 
$ \left\langle e_1|e_1'\right\rangle_o 
= \left\langle T(e_1)|T(e_1')\right\rangle_o 
= \left\langle e_2|e_2'\right\rangle_o$ 
with the $Aut(A)$-invariant inner product from Lemma~6. So we get a real number $s_o$ with 
$\left\langle e,e'\right\rangle_o = s_o$ for all atoms $e$.

With the Cauchy-Schwarz inequality we get
$ \left|\left\langle e,e'\right\rangle_o\right| 
\leq \left\langle e,e \right\rangle_o \left\langle e',e' \right\rangle_o
= 1$
and 
$\left|\left\langle e,e' \right\rangle_o\right| = 1$ is impossible,
since $e$ and $e'$ are linearly independent.
Thus $\left|s_o\right| < 1$, what we now use to define a further inner product 
on $A$ (see \cite{muller2012ududec} for a similar construction):
$$ \left\langle a|b\right\rangle_1 
:= \frac{1}{1 - s_o} \left[ \left\langle a|b\right\rangle_o 
- \frac{s_o}{(1+s_o)^{2}} \left\langle \mathbb{I}|a \right\rangle_o \left\langle \mathbb{I}|b \right\rangle_o \right] $$
for $a,b \in A$. For the atoms $e$ we have 
$ \left\langle \mathbb{I}|e\right\rangle_o 
= \left\langle e + e'|e\right\rangle_o = 1 + s_o$
and therefore $\left\langle e|e\right\rangle_1 = 1$ and 
$\left\langle e|e'\right\rangle_1 = 0$.
Furthermore, any $a \in A$ has the form $a = se + te'$ with an atom $e$ and $s,t \in \mathbb{R}$ 
and
$ \left\langle a|a \right\rangle_1 =  \left\langle se + te' | se + te'\right\rangle_1 = s^{2} + t^{2}$.
Thus $\left\langle a|a \right\rangle_1 \geq 0$ and
$\left\langle a|a \right\rangle_1 = 0$ iff $a=0$.

With $V := \left\{v \in A | \left\langle \mathbb{I}|v \right\rangle_1 = 0 \right\}$
we have $A = \mathbb{R} \mathbb{I} \oplus V$. For $v,w \in V$ and $s,t \in \mathbb{R}$ 
we define the following bilinear commutative product on $A$:
$$ (v + s \mathbb{I}) \circ (w + t \mathbb{I}) 
= tv + sw + ( \frac{1}{4} \left\langle v|w\right\rangle_1  + st) \mathbb{I}.$$
With this product $A$ becomes a spin factor (\cite{AS02} 3.33)
and spin factors are Jordan algebras (\cite{AS02} 3.37).
Note that any positive multiple of an inner product provides another one
and the parameter $1/4$ is conveniently chosen here 
to make sure that the atoms become idempotent elements,
what we will show now.

Obviously $\mathbb{I}$ is the identity for this product.
If $e$ is an atom, then $e'$ is an atom because of $m=2$ and
$\left\langle \mathbb{I}|e - e'\right\rangle_1 
= \left\langle e + e'|e - e'\right\rangle_1 
= \left\langle e|e\right\rangle_1 - \left\langle e'|e'\right\rangle_1 = 0$.
This means $e - e' \in V$. Furthermore 
$ \left\langle e - e' | e - e'\right\rangle_1 = 2$, 
$e = (e-e')/2 + \mathbb{I}/2$ and 
$e^{2} = e\circ e = (e - e')/2 + (\left\langle e-e'|e-e'\right\rangle_1/8 + 1/4) \  \mathbb{I}
= (e - e')/2 + 1/4 \mathbb{I}= e$.

Moreover $e \circ e' = e \circ (\mathbb{I} - e) = e - e^{2} = 0$ and
$(s e + t e')^{n} = s^{n} e + t^{n} e' $ for any $s,t \in \mathbb{R}$, $n \in \mathbb{N}$.
Thus we have $a^{2} \geq 0$ for any $a \in A$, $a^{2} = 0$ iff $a = 0$
and each positive element in $A$ is the square of some other element. 

(ii) Suppose $m \geq 4$, the weak symmetry and the gbit-property are given
and the atoms form a connected set.
We know from Lemma 1 that $L_A = ext\left[0,\mathbb{I}\right]$ 
is an orthomodular lattice, which becomes atomic and complete since the information capacity is finite. 
By Lemma 5, $L_A$ possesses the covering property and by Lemma 4, $L_A$ is irreducible.

Every irreducible complete orthomodular lattice with the covering property and $m \geq 4$
is isomorphic to the lattice of orthoclosed subspaces of a m-dimensional 
inner product space over some division ring 
\cite{piron1964axiomatique}. This division ring can be rather exotic \cite{Keller1980}. 
MacLaren \cite{maclaren1965notes} identified 
the following two conditions that make the division ring become the real numbers, complex numbers 
or quaternions:
\begin{itemize}
\item[(1)]
The set of all atoms $e$ with $e \leq p$ is compact for any $0 \neq p \in L_A$.
\item[(2)]
For some $p \in L_A$ and real interval $I$, there exists a continuous nonconstant function 
$t \rightarrow q_t$ from $I$ to $\left\{q \in L_A | q \leq p \right\}$.
\end{itemize}
We first show (1). Select any atom $e_o$. The set of all atoms is identical 
with $\left\{T(e_o) | T \in Aut(A) \right\}$ and becomes compact therefore.
For any  $0 \neq p \in L_A$, the intersection of this set with the closed set 
$\left\{a \in A | 0 \leq a \leq p\right\}$ is then compact as well and we have (1).

Condition (2) is fulfilled with $p:=\mathbb{I}$, $I := \left[0,1\right]$,
any atoms $e_1 \neq e_2$ and a continuous path $q_t, t \in \left[0,1\right]$ 
from $e_1$ to $e_2$. 

Hence $L_A$ is isomorphic to the lattice of closed subspaces of the m-dimensional
Hilbert space over the real numbers, complex numbers or quaternions. This lattice 
is isomorphic to the projection lattice $L_{H_m} = \left\{ p \in H_m | p = p^{2} \right\}$ 
in the Jordan matrix algebra $H_m$ over the division ring. Now let 
$\pi : L_{H_m} \rightarrow L_A$ denote the isomorphism. 

With the unique spectral representation of the elements in $H_m$ as well as in $A$
we can easily extend $\pi$ to a bijection between $H_m$ and $A$ that preserves
positivity and the norm, but we do not get the linearity of the extension that way.
We shall now tackle this problem using a generalized version of Gleason's theorem 
for Jordan matrix algebras (\cite{bunce1985Jordan} Theorem 2.2) and following
the lines of the proof of \cite{bunce1992mackey} Lemma 1.1.

Each linear functional $ \rho $ on the order unit space $A$ can be written 
as the difference of two positive functionals $\mu$ and $\nu$
(see for instance \cite{hanche1984jordan} Lemma 1.2.6). Then 
$\mu \pi$ and $\nu \pi$ become non-negative measures on $L_{H_m}$ and possess
linear extensions to $H_m$ by the generalized version of Gleason's theorem. 
Therefore there is a linear extension $\tilde{\rho}$ of $ \rho \pi$ to $H_m$ 
for each $\rho$ in the dual space $A^{\ast}$.

Now let $ x = \sum s_k p_k $, $s_k \in \mathbb{R} $, be any linear combination 
of elements in $L_{H_m}$ and $\rho \in A^{\ast}$. Then
\begin{center}
$\rho(\sum s_k \pi(p_k)) = \sum s_k \rho (\pi(p_k))
= \tilde{\rho} (\sum s_k p_k) 
= \tilde{\rho}(x)$
\end{center}
and, using the inequality from section 2 for the norm of $\tilde{\rho}$,
\newpage
\begin{align*}
\left|\rho(\sum s_k \pi(p_k)) \right|
&= \left|\tilde{\rho}(x)\right| 
\leq \left\| x \right\| \! \left\| \tilde{\rho} \right\| 
\leq 2 \left\| x \right\| sup \left\{ \left|\tilde{\rho}(p)\right| : p \in L_{H_m}  \right\} \\
&= 2 \left\| x \right\| sup \left\{ \left|\rho \pi (p)\right| : p \in L_{H_m} \right\}
= 2 \left\| x \right\| sup \left\{ \left|\rho (q)\right| : q \in L_A \right\} \\
&\leq 2 \left\|x\right\| \! \left\|\rho\right\|.
\end{align*}
Since this holds for all $\rho \in A^{\ast}$, we have
$$ \left\| \sum s_k \pi(p_k) \right\| \leq 2 \left\| x\right\| = 2 \left\|\sum s_k p_k \right\|$$
and particularly
$\sum s_k \pi(p_k) = 0$ for $\sum s_k p_k = 0$.
Therefore $\tilde{\pi}(x) := \sum s_k \pi(p_k)$ becomes well-defined 
for $x = \sum s_k p_k$ and we get a linear extension
$\tilde{\pi}$ of $\pi$ to $H_m$.
This extension is unique. Since $A$ is the linear span of $L_A$, we get that  
$\tilde{\pi} : H_m \rightarrow A$ is surjective and the spectrality of $H_m$ 
gives us the injectivity (note that orthogonal elements in $L_A$ 
are linearly independent in$A$). 
Moreover, both $\tilde{\pi}$ and its inverse 
are positive, which again follows from the spectrality of $A$ and $L_{H_m}$
(each $a$ with $0 \leq a$ has a spectral representation with non-negative coefficients).
Thus $A$ with the product 
$a \circ b := \tilde{\pi} \left[ \tilde{\pi}^{-1}(a) \circ \tilde{\pi}^{-1}(b) \right]$
for $a,b \in A$ becomes a Jordan algebra which is isomorphic to $H_m$.\hfill $\square$\\

MacLaren's result \cite{maclaren1965notes} bases upon earlier work 
\cite{maclaren1964atomic, weiss1958locally, zierler1961axioms, zierler1966lattice} 
and ultimately on Pontrjagin's theorem which states that there are 
only three (topologically) connected division rings, namely
the real numbers, complex numbers and the quaternions~\cite{pontrjagin1932stetige}.

Many authors have contributed to step-by-step 
extensions of Gleason's theorem from 1957 \cite{gleason1957} 
up to the recent most general version \cite{Gleason-vector-valued}
for the vector-valued measures on Jordan algebras (see \cite{Gleason-vector-valued} 
for the historical details). We could have used this new version, but can well do without,
since the positive linear functionals on an order unit space generate the dual.\\

\textbf{Corollary 1:} \itshape Suppose that $A$ is a spectral order unit space with
finite dimension, information capacity $m \neq 3$ and 
a strongly order determining state space. If the gbit-property holds and
$A$ is continuously symmetric, then $A$ is a 
simple (irreducible) Euclidean Jordan algebra.\\
\normalfont

This corollary follows immediately from Theorem 1, since the continuous symmetry
includes the weak symmetry and implicates that the atoms form a connected set.

The proof of our results is extremely different in the two cases
with information capacity $m = 2$ and $m \geq 4$.
An alternative proof is available for the first case 
\cite{BarnumHilgert2020, Dakic_Bru2011}.

The case $m=3$ remains a little enigmatic.
Piron's representation theorem for the  orthomodular lattices \cite{piron1964axiomatique} 
is not applicable then.
The Hermitian $3 \times 3$-matrices over the real numbers, complex numbers, quaternions
and in this case also over the octonions form Euclidian Jordan algebras with $m=3$
and satisfy all our postulates, but the existence of further irreducible examples 
with $m=3$ remains unclear here.

\section{Conclusions}

Any physical theory needs mathematical models for the dynamical processes. When we assume 
that these processes are reversible and continuous in time
and that the atoms can continuously evolve from one to each other, 
we come to the symmetry 
postulate that we have studied here. It rules out the classical 
and the reducible finite-dimensional cases and brings us to the 
simple (irreducible) Euclidean Jordan algebras
(spin factors and Jordan matrix algebras over the real numbers, complex numbers and quaternions). 
It thus becomes a far-reaching basic principle for the reconstruction of quantum theory.

The question why nature in quantum theory prefers the complex numbers over the real numbers 
and the quaternions is not considered here, but
has already been addressed in different ways. One way is to consider locally tomographic models 
for composite systems \cite{barnum2014local, barrett2007information, Chiribella-PRA2011, 
delatorre2012PRL-from_local_and_rev, masanes2011derivation, nie2020loc_tomography}. 
Other ways are to postulate somehow that an observable plays
the role of the Hamiltonian (dynamcial correspondence \cite{AS02, nie2015dyn}, 
energy observability \cite{barnum2014higher}).

The case with information capacity $m=3$ is somewhat exceptional and remains a little enigmatic.
Furthermore, many examples are known with $m=2$ and infinite dimension (for instance the 
spin factors with infinite dimension), but it is not evident whether 
the case $3 \leq m < \infty$ includes infinite-dimensional order unit spaces.
We had to presuppose the finite dimension to get a compact automorphism group.

We have needed the \emph{gbit-property}, which turned out to be 
the \emph{covering property} in disguise. The physical or information theoretical necessity 
of these properties might be disputable and it is not clear
what mathematical structure would arise when we do without them.

We have not needed any one of the following postulates 
that are part of various reconstructions of quantum theory:
\emph{symmetry of the transition probability} \cite{nie2020charJordan},
\emph{bit symmetry} \cite{muller2012ududec},
\emph{strong symmetry} \cite{barnum2014higher},
absence of \emph{third-order interference} \cite{barnum2014higher, nie2020charJordan}
and the existence of \emph{compressions} or \emph{filters} \cite{AS02, AIHPA1978Guz, guz1980conditional, mielnik1969theory}.
These postulates are satisfied by the Jordan matrix algebras 
and thus become deducible features in our approach.

The compressions are an abstract mathematical notion with an important physical meaning; they represent 
an axiomization of the quantum measurement process \cite{AS02}. In the approach presented here their existence 
is a consequence of the continuous symmetry.
\bibliographystyle{abbrv}
\bibliography{Literatur2024}
\end{document}